\documentclass[sigconf,nonacm]{acmart}
\usepackage{tabularx}
\usepackage[htt]{hyphenat}
\AtBeginDocument{%
  \providecommand\BibTeX{{%
    \normalfont B\kern-0.5em{\scshape i\kern-0.25em b}\kern-0.8em\TeX}}}

\setcopyright{none}
\renewcommand\footnotetextcopyrightpermission{}
\copyrightyear{2023}
\acmYear{2023}
\acmDOI{XXXXXXX.XXXXXXX}

\acmConference[AIDB]{5th International Workshop on Applied AI for Database Systems and Applications}{Sept 1,
  2023}{Vancouver, Canada}
\acmPrice{15.00}
\acmISBN{978-1-4503-XXXX-X/18/06}

\begin{document}

\title{Comprehending Semantic Types in JSON Data with Graph Neural Networks}


\author{Shaung Wei}
\email{sw2582@rit.edu}
\affiliation{%
  \institution{Rochester Institute of Technology}
  \streetaddress{102 Lomb Memorial Drive}
  \city{Rochester}
  \state{New York}
  \country{USA}
  \postcode{14623-5608}
}

\author{Michael J. Mior}
\email{mmior@mail.rit.edu}
\affiliation{%
  \institution{Rochester Institute of Technology}
  \streetaddress{102 Lomb Memorial Drive}
  \city{Rochester}
  \state{New York}
  \country{USA}
  \postcode{14623-5608}
}


\begin{abstract}
Semantic types are a more powerful and detailed way of describing data than atomic types such as strings or integers. They establish connections between columns and concepts from the real world, providing more nuanced and fine-grained information that can be useful for tasks such as automated data cleaning, schema matching, and data discovery. Existing deep learning models trained on large text corpora have been successful at performing single-column semantic type prediction for relational data. However, in this work, we propose an extension of the semantic type prediction problem to JSON data, labeling the types based on JSON Paths. Similar to columns in relational data, JSON Path is a query language that enables the navigation of complex JSON data structures by specifying the location and content of the elements. We use a graph neural network to comprehend the structural information within collections of JSON documents. Our model outperforms a state-of-the-art existing model in several cases. These results demonstrate the ability of our model to understand complex JSON data and its potential usage for JSON-related data processing tasks.
\end{abstract}


\maketitle

\section{Introduction}
Detecting the semantic types of columns in a relational table can be useful for data preparation and information extraction tasks such as data cleaning and integration. For example, semantic type detection can help some rule-based automated data cleaning applications that are dependent on semantic data types~\cite{kandel2011wrangler,raman2001potter}. Schema matching tasks can also narrow down search spaces based on detected semantic types~\cite{rahm2001survey}. Furthermore, data discovery tasks can utilize detected semantic types to find semantically related results for input queries~\cite{fernandez2018aurum,fernandez2018seeping}.

Traditional systems only classify table columns into atomic types such as Boolean, integer, and string. Semantic types are a finer-grained classification of columns that provide richer information and a connection to real-world concepts. For example, a column containing values such as "Chicago", "Detroit" can be described using the type \emph{location} rather than \emph{string}. Prior work has attempted to predict semantic types using methods such as dictionary lookup and regular expression matching. In real-world cases, many tables are dirty with corrupted column names or missing values that these rule-based approaches do not correctly predict~\cite{mind}.

To solve this problem, a deep learning enabled model, Sherlock~\cite{hulsebos2019sherlock}, was proposed to learn the semantic type based on column values. Sherlock is trained on huge table-based corpora~\cite{hu2019viznet}. It first extracts features from the values in each column and a deep learning model is trained on these features to perform semantic type detection. Although Sherlock outperforms traditional approaches, it is limited to relational data.

In this work, we propose a novel semantic type classification model for semi-structured JSON data. Unlike relational databases, JSON data is in the form of key-value pairs and uses a hierarchical structure. We annotate the semantic type of JSON data with its hierarchical structure and use a graph neural network to predict the semantic type with the same set of features extracted by Sherlock. The proposed model can achieve better accuracy and higher F1 scores than Sherlock on certain semantic types.

\section{Problem Setup}

Relational semantic type prediction is a multiclass classification problem that can be defined as follows. Given the columns $c_1,c_2,\ldots,\\c_m$ for a given table as the training dataset, where $c_i$ is a vector of column values, their corresponding semantic types $t_1,t_2,\ldots,t_m \in \tau$ are considered as target values, where $\tau$ is a set of predefined semantic labels to be considered. We refer to this problem as \textit{single-column prediction} where we use all values from each column to predict the semantic type of the column. 

\begin{figure}
  \centering
  \includegraphics[width=0.4\textwidth]{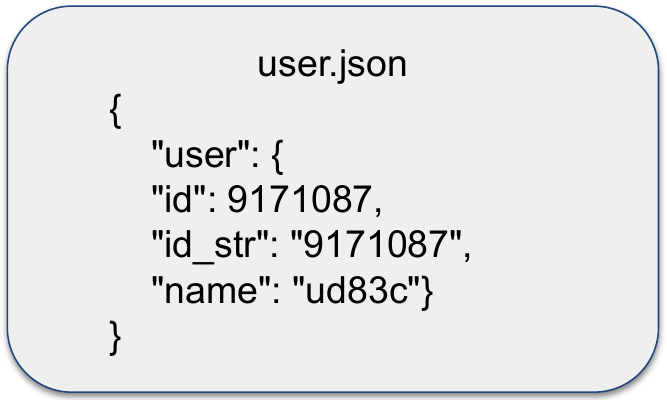}
  \caption{Example of JSON Data}
  \label{fig:json_ex}
\end{figure}

In our work, we extend this problem to include JSON data, which contains arbitrarily nested structures. To label JSON data, an example JSON document shown in Figure~\ref{fig:json_ex} could include a \texttt{user} key with the following key-value pairs: \texttt{"user": \{"id":9171087, "id\_str": "9171087", "name": ud83c\}}. In this instance, the JSON document yields four distinct key-value pairs. The first key-value pair consists of the key \texttt{id} and its corresponding value of \texttt{9171087}. The second pair includes the key \texttt{id\_str} with its corresponding value also being \texttt{9171087}. The third key-value pair comprises the key \texttt{name} with its corresponding value being \texttt{ud83c}. The fourth pair, has the key \texttt{user} and the value \texttt{\{"id":9171087, "id\_str": "9171087", and "name": ud83c\}}.

In the context of relational data, the assignment of ground truth values for supervised semantic type prediction is often derived from columns, as they are indicative of the underlying meaning of the data within that column. In the case of JSON data, the semantic type is established by reference to the corresponding JSON Path\cite{jsonpath} of the key. JSON Path is a query language that enables the navigation of complex JSON data structures by specifying the element location. The rationale is that values associated with the same path are likely to possess similar semantic meanings, and therefore can be grouped together under a common semantic type. For example, the path \texttt{\$.user.username} refers to the "username" key within the top-level object nested under the key "user".

\section{Feature Selection}
In our work, we use the same set of features as Sherlock, which is a single-column prediction model that takes all values from a single column as input and outputs the predicted semantic type for the corresponding column. The extracted feature vectors are used to train a neural network for the detection of semantic types. In Sherlock, a total of 1,587 features are extracted from the column values across four different dimensions. These dimensions are described below.
\begin{enumerate}
  \item \textbf{Global statistics.} This category is a set of 27 hand-crafted features, typically some high-level statistical characteristics of the column. For example, column entropy describes the uniformity of the distribution of column values. Another example is the number of values that measures the number of unique values recorded in the column.
  \item \textbf{Character-level distributions.} This category contains simple statistical features of character distributions. Specifically, 10 statistical functions, i.e. any, all, mean, variance, min, max, median, sum, kurtosis, skewness, are applied on all 96 ASCII-printable characters plus the form feed character, resulting in 960 features. For example, the \texttt{any} function checks if any column value contains a specific character. \texttt{all} checks if all column values contain a character. Some other examples of features are the maximum number of appearances of a character in a single column value and the median value of appearance of a character for all column values.
  \item \textbf{Word embeddings.} Sherlock uses a pre-trained GloVe embedding~\cite{pennington2014glove} to characterize the semantic content of column values. The GloVe model contains a 50-dimensional embedding for 400,000 English words aggregated from 6,000,000,000 tokens. Similar to word2vec~\cite{mikolov2013efficient}, GloVe embeddings can be used to measure semantic similarity between words. The advantage of GloVe compared to word2vec is that it does not rely simply on local information of words, but it also incorporates global statistics such as word co-occurrence. By calculating the mean, mode, median, and variance on the 50-dimensional GloVe feature vector for all column values, a 200-dimensional feature vector is produced in this category.
  \item \textbf{Paragraph vectors.} A distributed bag of words version of the paragraph vector (PV-DBOW), or the doc2vec model~\cite{pmlr-v32-le14} is implemented to capture features at the ``topic'' level of the column. The doc2vec model forces the model to predict random words that are sampled from paragraphs in the output by ignoring context words from the input. The model is pre-trained in Sherlock using the Gensim library to extract a paragraph feature that has 400 dimensions. 
\end{enumerate}

\section{Proposed Graph Model}

Raw JSON data is annotated by treating each key-value pair as a data point. For instance, from the \texttt{user.json} file, four data points can be extracted as illustrated in Figure~\ref{fig:json_ex}. For each key-value pair, the label is determined by annotating the JSON Path to represent the semantic meaning, while the features are extracted using Sherlock from the corresponding values at each path. As an example, consider the key-value pair \texttt{"user": \{"id":9171087, "id\_str": "9171087", "name": "ud83c"\}}. Here, the label assigned is \texttt{user}, and the features are extracted from \texttt{\{"id":9171087, "id\_str": "9171087", "name": "ud83c"\}}. At this stage, we can proceed to apply Sherlock and evaluate its performance for semantic type detection, using it as a baseline result.

\begin{figure}
  \centering
  \includegraphics[width=0.5\textwidth]{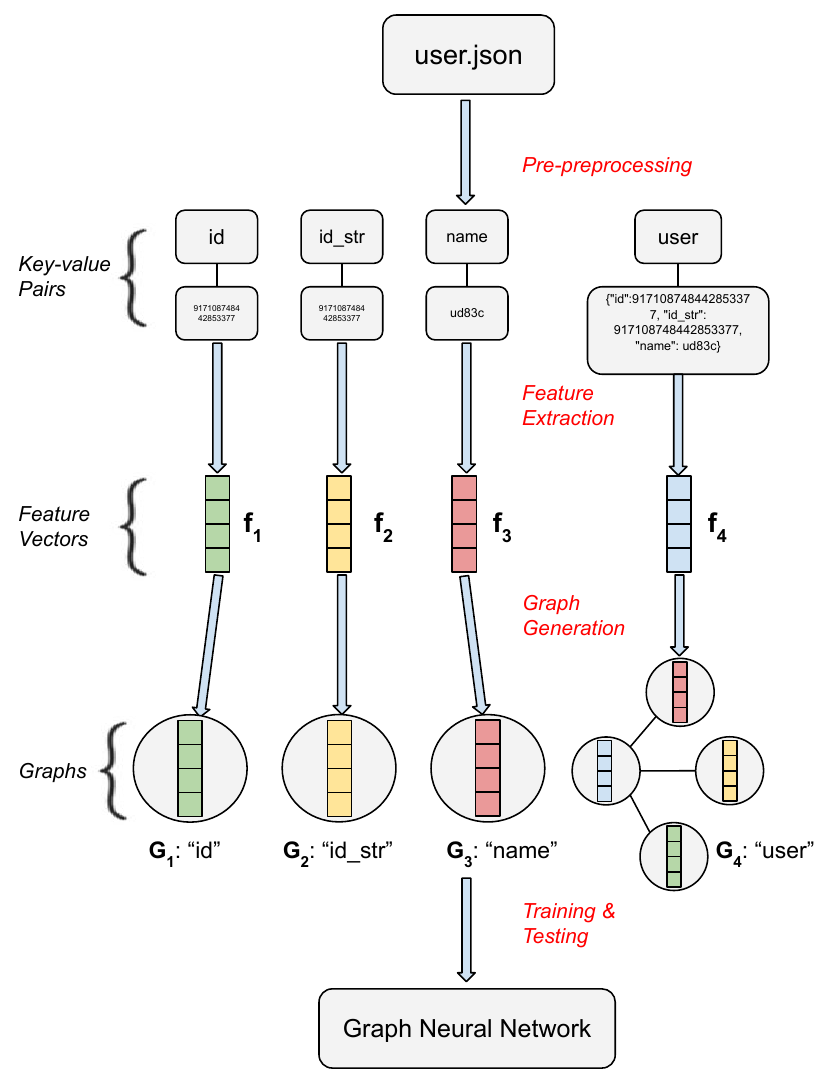}
  \caption{Proposed model architecture}
  \label{fig:json_to_graph}
\end{figure}

The structure of our model is illustrated in Figure~\ref{fig:json_to_graph}. Each JSON file is processed by first obtaining all key-value pairs, as described in the preceding section. Using the same example as in Figure~\ref{fig:json_ex}, we can obtain four key-value pairs. Subsequently, the features for each key-value pair are computed based on their respective values, yielding features $f_1$, $f_2$, $f_3$, and $f_4$.

Following this, four graphs are generated, labeled as "id", "id\_str", "name", and "user". The first three graphs, $G_1$, $G_2$, and $G_3$, each consist of a single-node, with the node features $f_1$, $f_2$, and $f_3$. The fourth graph, $G_4$, is a multi-node graph, with a root node having node features $f_4$ and three edges connecting to three other nodes, each having node features $f_1$, $f_2$, and $f_3$. 

Once we obtain the graph representations, we use graph neural networks (GNNs)\cite{gcn} to perform the classification task. Graph neural networks are well-suited for our problem, as JSON documents inherently possess a tree-like structure that can be represented as a graph. GNNs excel at capturing complex dependencies and relationships within graphs by leveraging structural information encoded in edges and node features. Using GNNs, we can effectively leverage the hierarchical relationships and dependencies present in JSON documents, enabling accurate analysis and prediction tasks such as semantic type prediction or classification. GNNs exploit the rich structural information of JSON documents, making them a powerful approach to understanding and extracting insights from this graph-based representation. In our study, we used the Spektral library\cite{spectral} to implement our GNN. Specifically, we employed a two-layer GCN model, where the input graph was first fed into a GCN layer with 256 hidden units, followed by graph pooling and a dropout layer. The output was then forwarded to another GCN layer with 64 hidden units, before being connected to a dense layer for multi-class classification. To optimize the model, we adopted Adam\cite{adam} as our optimizer and used categorical cross-entropy as our loss function. The learning rate was set to $2 \times 10^{-4}$.

\section{Dataset}
Our study uses Twitter data and Meetup data available on push-shift.io~\cite{pushdata}, a large-scale archive of social media content. Due to the immense size of the dataset, we selected a representative subset of each dataset for our research. Specifically, we extracted all available data from a specific date, resulting in a total of 30,000 distinct JSON objects for Twitter data, and 20,000 distinct JSON objects for Meetup data.

\begin{figure}
  \centering
  \includegraphics[width=0.4\textwidth]{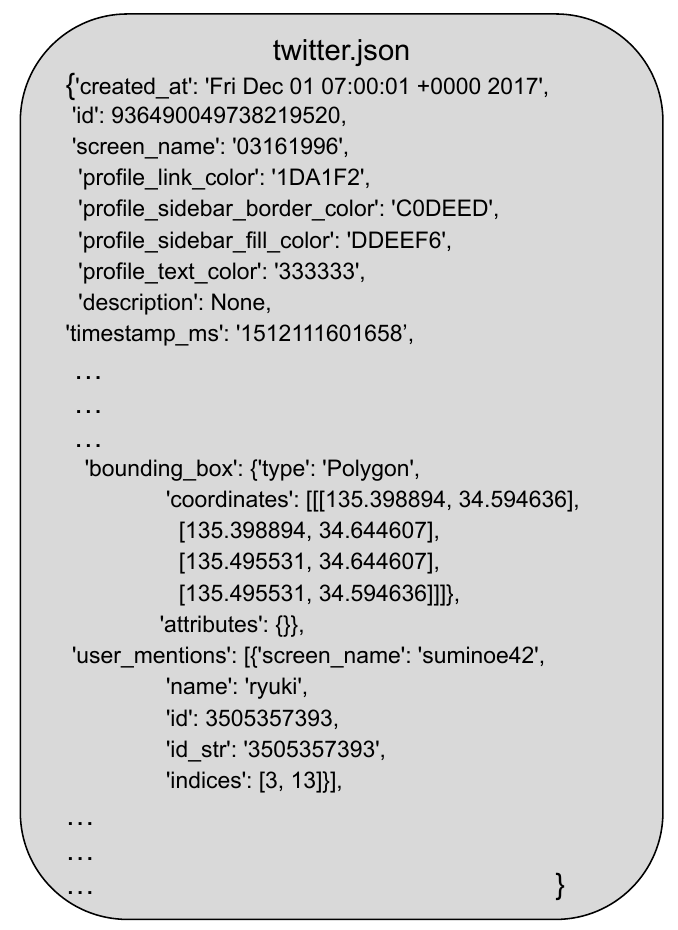}
  \caption{Example of Twitter JSON data}
  \label{fig:twitter_ex}
\end{figure}

Figure~\ref{fig:twitter_ex} shows an example of the Twitter dataset used in this study. The example contains labels for a single-node graph, including \texttt{created\_at}, \texttt{id}, and \texttt{screen\_name}. The \texttt{profile\_link\_color}, \texttt{profile\_sidebar\_border\_color}, \texttt{profile\_sidebar\_fill\_color}, and \texttt{profile\_text\_color} also belong to single-node graphs with label color. Additionally, the example comprises multi-node graphs with labels such as \texttt{bounding\_box} and \texttt{user\_mentions}, which are also illustrated in the figure. The above-mentioned nodes have a depth of 1 in the graph, while nodes such as \texttt{type} and \texttt{coordinate} in this example have a depth of 2. Table~\ref{tab:depth} presents the number of examples corresponding to each depth (level of nesting) for our two datasets.

\begin{table}[ht]
    \centering
    \begin{tabular}{ccc}
        \hline
        Depth & \multicolumn{2}{c}{Number of examples} \\
        \cline{2-3}
        & Twitter & Meetup\\
        \hline
        1 & 1,205 & 191,884 \\
        2 & 32,158 & 304,750 \\
        3 & 74,187 & 101,930 \\
        4 & 3,907 & 0 \\
        5 & 124 & 0 \\
        \hline

    \end{tabular}
    \caption{Number of examples for each depth}
    \label{tab:depth}
\end{table}

All key-value pairs with a null value or a type of Boolean are ignored since these values are highly repetitive in our dataset and do not contain useful semantic information. We then annotated each JSON Path with a class label, as discussed in the previous section, resulting in a label set comprising 43 distinct classes for the Twitter dataset and 32 distinct classes for the Meetup dataset. 

To prepare the data for use in our model, we processed each JSON file into a graph structure, resulting in a total of around 110,000 distinct graphs for the Twitter dataset and 600,000 distinct graphs for the Meetup dataset. Each graph in our dataset is accompanied by a set of node features, an adjacency matrix, and a class label encoded in one-hot format. The dataset is partitioned into training, validation, and test sets in a 7:3:3 ratio.

\section{Experiment Method}
\begin{figure*}
  \centering
  \includegraphics[width=\textwidth]{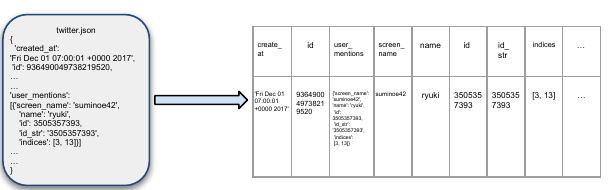}
  \caption{transformation of JSON document to relational table}
  \label{fig:json_to_tab}
\end{figure*}

In our experiment, we initially preprocess the JSON file by extracting key-value pairs based on their corresponding JSON paths. Figure~\ref{fig:json_to_tab} demonstrates the conversion of a JSON document into relational tables, where each key-value pair is represented as a separate column. Subsequently, a feature extraction process is applied to these values to obtain the feature vectors. This step enables us to utilize the Sherlock model for classification, allowing us to establish a baseline performance measure. We then proceed to process the data into graphs, following the procedures outlined in Figure~\ref{fig:json_to_graph}. The classification task is then performed using a graph neural network model. We summarize the time spent on each preprocessing step in Table~\ref{tab:preprocessing}, providing an overview of the time requirements for these tasks on our two different datasets. The key-value pair extraction step takes 1,205s and 3,525s for the Twitter and Meetup datasets respectively, which is the most time-consuming preprocessing step. The feature extraction and graph processing steps take less time than the key-value pair extraction. The feature extraction step is the only preprocessing step required to train the Sherlock model. 

\begin{table}[ht]
    \centering
    \begin{tabular}{ccc}
        \hline
        Preprocessing steps & Twitter & Meetup \\
        \hline
        Key-value pair extraction & 1,205s & 3,525s \\
        Feature extraction & 785s & 2,100s \\
        Graph processing & 152s & 450s \\
        \hline
        Total time & 2,142s & 7,075s \\
        \hline

    \end{tabular}
    \caption{Time spent on preprocessing steps}
    \label{tab:preprocessing}
\end{table}

\section{Experiment Results}
\begin{table}[htbp]
  \centering
  \caption{Comparison of Sherlock and the proposed model on Twitter dataset}
  \label{tab:comparison}
  \begin{tabularx}{\linewidth}{@{}l *{5}{>{\centering\arraybackslash}X}@{}}
    \hline
     &  & \multicolumn{2}{c}{Sherlock} & \multicolumn{2}{c}{Proposed model} \\
    \cline{3-4} \cline{5-6}
    & Label & \small{F1 score} & \small{Accuracy} & \small{F1 score} & \small{Accuracy} \\
    \hline
    \small{Single-Node} & \small{screen\_name} & 0.92 & 0.93 & \textbf{0.95} & 0.93 \\
       & \small{country\_code} & 0.80 & \textbf{1.00} & \textbf{0.92} & 0.85 \\
       & \small{timestamp\_ms} & \textbf{1.00} & \textbf{1.00} & 0.97 & 0.96 \\
       & \small{color} & \textbf{0.99} & 0.99 & 0.98 & 0.99 \\
       & \small{description} & \textbf{0.75} & \textbf{0.77} & 0.67 & 0.60 \\
    \hline
    \small{Multi-Node} & \small{bounding\_box} & 0.83 & 1.00 & \textbf{1.00} & 1.00 \\
    & \small{user\_mentions} & 0.59 & 0.41 & \textbf{0.84} & \textbf{0.97} \\
    & \small{retweet\_status} & 0.57 & 0.48 & \textbf{0.82} & \textbf{0.86}\\
    & \small{hashtag} & 0.34 & 0.23 & \textbf{0.40} & \textbf{0.80} \\
    & \small{full\_name} & 0.00 & 0.00 & \textbf{0.22} & \textbf{0.97} \\
    \hline
    Average &  & 0.82 & 0.84 & \textbf{0.85} & \textbf{0.85} \\
    \hline
  \end{tabularx}
\end{table}

Table~\ref{tab:comparison} provides a comprehensive comparison between the performance of Sherlock and the proposed model in terms of the F1 score and accuracy metrics. The evaluation is conducted for all semantic classes, which are categorized into single-node and multi-node based on the number of nodes in the graph. The table presents some selected examples from each category, including \texttt{screen\_name}, \texttt{country\_code}, \texttt{timestamp\_ms}, color, and description for single-node, and \texttt{bounding\_box}, \texttt{user\_mentions}, \texttt{retweet\_status}, \texttt{hashtag}, and \texttt{full\_name} for multi-node. The average performance result is also presented at the bottom of the table.

In the single-node setting, the proposed model outperforms Sherlock on some labels, for example, with F1 scores of 0.95 for \texttt{screen\_name}, 0.92 for \texttt{country\_code}, compared to 0.92 and 0.80, respectively, for Sherlock. However, Sherlock achieves a perfect F1 score of 1.00 for the \texttt{timestamp\_ms} label, while the proposed model only achieves a score of 0.97. The Sherlock model also exhibits higher accuracy levels for the semantic types of \texttt{country\_code} and \texttt{timestamp\_ms}. Sherlock also performs slightly better for the \texttt{description} label, with an F1 score of 0.75 compared to 0.67 for the proposed model. Based on the results, it can be suggested that the base Sherlock model might exhibit better performance when applied to single-node scenarios, which may be attributed to the fact that the Sherlock model utilizes a more complex neural network architecture compared to our network.

In the multi-node setting, the proposed model achieves a perfect F1 score of 1.00 for the \texttt{bounding\_box} label, while Sherlock achieves a score of 0.83. The proposed model also outperforms Sherlock for the \texttt{user\_mentions} and \texttt{retweet\_status} labels, with F1 scores of 0.84 and 0.82 compared to 0.59 and 0.57, respectively. However, Sherlock achieves a slightly better F1 score of 0.40 for the hashtag label compared to 0.34 for the proposed model. The proposed model achieves a higher F1 score of 0.22 for the \texttt{full\_name} label compared to 0.00 for Sherlock. In terms of accuracy, our proposed model significantly outperforms Sherlock. Specifically, for the hashtag semantic type, our model achieves an accuracy of 0.80, while Sherlock only achieves an accuracy of 0.23. Similarly, for the \texttt{full\_name} semantic type, our model achieves an accuracy of 0.50, while Sherlock fails to classify any instance correctly. These findings demonstrate that our proposed model is more adept at predicting complex structures, thus providing greater utility for practical applications.

Table~\ref{tab:meetupcomp} presents the comparison between Sherlock and our proposed model on the meetup dataset. The JSON files within the meetup dataset exhibit highly similar structures, resulting in higher overall prediction accuracy and F1 score. In cases involving multiple nodes, our model achieves perfect accuracy of 1.00, while Sherlock performs equally well on classes such as event and category, with F1 scores of 0.99 and 0.97 respectively for \texttt{group} and \texttt{group\_photo}. For single-node classes, both models demonstrate similar performance levels. The Meetup dataset consists of numerous homogeneous JSON files, many of which exhibit similar hierarchical structures. As a result, the base model Sherlock can achieve good performance in the multi-node setup, indicating that our model does not outperform Sherlock on this dataset. The reason could be that the Meetup dataset exhibits a higher level of homogeneity and has a less complex hierarchical structure for our model to take advantage of.

\begin{table}[htbp]
  \centering
  \caption{Comparison of Sherlock and the proposed model on the Meetup dataset}
  \label{tab:meetupcomp}
  \begin{tabularx}{\linewidth}{@{}l *{5}{>{\centering\arraybackslash}X}@{}}
    \hline
     &  & \multicolumn{2}{c}{Sherlock} & \multicolumn{2}{c}{Proposed model} \\
    \cline{3-4} \cline{5-6}
    & Label & \small{F1 score} & \small{Accuracy} & \small{F1 score} & \small{Accuracy} \\
    \hline
    \small{Single-Node} & \small{ event\_id} & 0.66 & \textbf{0.64} & \textbf{0.68} & 0.63 \\
       & \small{id} & 0.86 & 0.85 & \textbf{0.91} & \textbf{0.90} \\
       & \small{member\_name} & \textbf{0.95} & \textbf{0.95} & \textbf{0.95} & 0.94 \\
       & \small{shortname} & \textbf{0.99} & 0.99 & 0.98 & \textbf{1.00} \\
    \hline
    \small{Multi-Node} 
    & \small{event} & \textbf{1.00} & \textbf{1.00} & \textbf{1.00} & \textbf{1.00} \\
    & \small{category} & \textbf{1.00} & \textbf{1.00} & \textbf{1.00} & \textbf{1.00}\\
    & \small{group} & 0.99 & 0.97 & \textbf{1.00} & \textbf{1.00} \\
    & \small{group\_photo } & 0.97 & 0.99 & \textbf{1.00} & \textbf{1.00} \\
    \hline
    Average &  & 0.89 & 0.89 & \textbf{0.92} & \textbf{0.90} \\
    \hline
  \end{tabularx}
\end{table}

\begin{table}
\centering
\caption{Training time and model size comparison}
\label{tab:traintime_modelsize}
\begin{tabular}{cccc}
\hline
Model & \multicolumn{2}{c}{Training time} & Model size \\
\cline{2-3}
& Twitter & Meetup &\\
\hline
Sherlock & 252s & 690s & 5.9MB \\
Proposed model & 1,230s & 3,051s & 1.9MB \\
\hline
\end{tabular}
\end{table}

Table \ref{tab:traintime_modelsize} shows a comparison between the training time and model size of Sherlock and the proposed model. Our proposed model takes significantly longer to train but produces a much smaller model. We expect that advances in graph neural network training will apply to our setting to further reduce the training time~\cite{bytegnn,scalablegnn}. Our proposed model exhibits a notable reduction in model size in comparison to Sherlock. This is mainly attributed to the fact that our model incorporates a smaller number of neural network layers. Consequently, this disparity may account for certain suboptimal predictions made by our model in comparison to Sherlock, particularly in single-node predictions.

Our proposed model achieves a higher average F1 score compared to Sherlock. These results suggest that our proposed model has a superior ability to learn structural information from the data.

\section{Future Work}
In our ongoing research, we aim to explore alternative subgraph representations to enhance the prediction of semantic types in our proposed model. The current subgraphs we utilize do not incorporate the parent node and its sibling nodes, thus missing out on valuable information present in those nodes. Sato~\cite{sato} is a model that utilizes neighbor columns and table-level information for semantic prediction; however, it exhibits limited scalability in terms of training time. To address this limitation, we plan to construct subgraphs that include the parent and sibling nodes of the target node. In addition, we will assign edge weights to the graph, allowing us to emphasize the importance of the target node for prediction. By enabling additional edge features within each subgraph, we can leverage neural networks such as edge-conditioned GCN~\cite{eccv} that are capable of utilizing edge weight information. Subsequently, we intend to combine the outputs obtained from different subgraphs using a transformer~\cite{transformer}. The transformer architecture is suitable for this purpose because of its ability to capture global dependencies and model the relationships between the outputs of different subgraphs effectively.

In addition, we plan to extend our experimentation beyond the current dataset in order to enhance the robustness and generalizability of our model. Specifically, we will train and test our model on additional datasets that are representative of real-world scenarios. Our goal is to evaluate the effectiveness of the model's predictive capabilities on JSON structures that it has not yet encountered. Moreover, we aim to examine the impact of training set size on the performance of our model, with particular attention to the detection of less frequent semantic types. By conducting such analyses, we aim to better understand the limitations and strengths of our model and further refine it for real-world applications.

\section{Conclusion}

In this paper, we proposed a model for predicting semantic types in nested JSON data that can be used for various automated data processing tasks. Existing models either predict for a single set of values at a time or are limited to non-nested relational data.

To address this limitation, we proposed an extension of the semantic type prediction problem to semi-structured JSON data with types labeled based on JSON Paths. Our proposed model annotates the semantic type of JSON data with its hierarchical structure and employs a graph neural network to predict the semantic type using the same set of features extracted by Sherlock. We demonstrate several cases where our model outperforms Sherlock, indicating its ability to comprehend complex JSON data and its potential for semi-structured data processing tasks.

Our ongoing research focuses on enhancing the prediction of semantic types in our proposed model through alternative subgraph representations. By incorporating the parent and sibling nodes into subgraphs and assigning edge weights, we can capture valuable information for accurate predictions. Leveraging additional edge features will further improve our model's performance. We plan to validate our approach on diverse datasets, ensuring its robustness and generalizability to real-world scenarios. Furthermore, we plan to investigate the impact of the size of the training set on model performance, especially for detecting less frequent semantic types.

In conclusion, our work contributes to the development of deep learning models for predicting semantic types in JSON data. Our proposed model offers a better understanding of complex data and improved performance compared to Sherlock on semi-structured data.
\balance

\bibliographystyle{ACM-Reference-Format}
\bibliography{references}


\begin{thebibliography}{21}


\ifx \showCODEN    \undefined \def \showCODEN     #1{\unskip}     \fi
\ifx \showDOI      \undefined \def \showDOI       #1{#1}\fi
\ifx \showISBNx    \undefined \def \showISBNx     #1{\unskip}     \fi
\ifx \showISBNxiii \undefined \def \showISBNxiii  #1{\unskip}     \fi
\ifx \showISSN     \undefined \def \showISSN      #1{\unskip}     \fi
\ifx \showLCCN     \undefined \def \showLCCN      #1{\unskip}     \fi
\ifx \shownote     \undefined \def \shownote      #1{#1}          \fi
\ifx \showarticletitle \undefined \def \showarticletitle #1{#1}   \fi
\ifx \showURL      \undefined \def \showURL       {\relax}        \fi
\providecommand\bibfield[2]{#2}
\providecommand\bibinfo[2]{#2}
\providecommand\natexlab[1]{#1}
\providecommand\showeprint[2][]{arXiv:#2}

\bibitem[Baumgartner(2021)]%
        {pushdata}
\bibfield{author}{\bibinfo{person}{Jason Baumgartner}.}
  \bibinfo{year}{2021}\natexlab{}.
\newblock \bibinfo{booktitle}{\emph{Push Shift Dataset}}.
\newblock
\urldef\tempurl%
\url{https://files.pushshift.io/}
\showURL{%
\tempurl}


\bibitem[Chen et~al\mbox{.}(2020)]%
        {scalablegnn}
\bibfield{author}{\bibinfo{person}{Ming Chen}, \bibinfo{person}{Zhewei Wei},
  \bibinfo{person}{Bolin Ding}, \bibinfo{person}{Yaliang Li},
  \bibinfo{person}{Ye Yuan}, \bibinfo{person}{Xiaoyong Du}, {and}
  \bibinfo{person}{Ji-Rong Wen}.} \bibinfo{year}{2020}\natexlab{}.
\newblock \showarticletitle{Scalable Graph Neural Networks via Bidirectional
  Propagation}. In \bibinfo{booktitle}{\emph{Advances in Neural Information
  Processing Systems}}, Vol.~\bibinfo{volume}{33}.
  \bibinfo{pages}{14556--14566}.
\newblock


\bibitem[Fernandez et~al\mbox{.}(2018a)]%
        {fernandez2018aurum}
\bibfield{author}{\bibinfo{person}{Raul~Castro Fernandez},
  \bibinfo{person}{Ziawasch Abedjan}, \bibinfo{person}{Famien Koko},
  \bibinfo{person}{Gina Yuan}, \bibinfo{person}{Samuel Madden}, {and}
  \bibinfo{person}{Michael Stonebraker}.} \bibinfo{year}{2018}\natexlab{a}.
\newblock \showarticletitle{Aurum: A data discovery system}. In
  \bibinfo{booktitle}{\emph{IEEE 34th International Conference on Data
  Engineering (ICDE)}}. \bibinfo{address}{Paris, France},
  \bibinfo{pages}{1001--1012}.
\newblock
\urldef\tempurl%
\url{https://doi.org/10.1109/ICDE.2018.00094}
\showDOI{\tempurl}


\bibitem[Fernandez et~al\mbox{.}(2018b)]%
        {fernandez2018seeping}
\bibfield{author}{\bibinfo{person}{Raul~Castro Fernandez},
  \bibinfo{person}{Essam Mansour}, \bibinfo{person}{Abdulhakim~A Qahtan},
  \bibinfo{person}{Ahmed Elmagarmid}, \bibinfo{person}{Ihab Ilyas},
  \bibinfo{person}{Samuel Madden}, \bibinfo{person}{Mourad Ouzzani},
  \bibinfo{person}{Michael Stonebraker}, {and} \bibinfo{person}{Nan Tang}.}
  \bibinfo{year}{2018}\natexlab{b}.
\newblock \showarticletitle{Seeping semantics: Linking datasets using word
  embeddings for data discovery}. In \bibinfo{booktitle}{\emph{IEEE 34th
  International Conference on Data Engineering (ICDE)}}.
  \bibinfo{address}{Paris, France}, \bibinfo{pages}{989--1000}.
\newblock
\urldef\tempurl%
\url{https://doi.org/10.1109/ICDE.2018.00093}
\showDOI{\tempurl}


\bibitem[Friesen(2019)]%
        {jsonpath}
\bibfield{author}{\bibinfo{person}{Jeff Friesen}.}
  \bibinfo{year}{2019}\natexlab{}.
\newblock \bibinfo{booktitle}{\emph{Extracting JSON Values with JsonPath}}.
\newblock \bibinfo{publisher}{Apress}, \bibinfo{address}{Berkeley, CA},
  \bibinfo{pages}{299--322}.
\newblock
\showISBNx{978-1-4842-4330-5}
\urldef\tempurl%
\url{https://doi.org/10.1007/978-1-4842-4330-5_10}
\showDOI{\tempurl}


\bibitem[Grattarola and Alippi(2021)]%
        {spectral}
\bibfield{author}{\bibinfo{person}{Daniele Grattarola} {and}
  \bibinfo{person}{Cesarei Alippi}.} \bibinfo{year}{2021}\natexlab{}.
\newblock \showarticletitle{Graph Neural Networks in TensorFlow and Keras with
  Spektral}.
\newblock \bibinfo{journal}{\emph{IEEE Computational Intelligence Magazine}}
  \bibinfo{volume}{16} (\bibinfo{year}{2021}), \bibinfo{pages}{99--106}.
\newblock
\urldef\tempurl%
\url{https://doi.org/10.48550/arXiv.2006.12138}
\showURL{%
\tempurl}


\bibitem[Hu et~al\mbox{.}(2019)]%
        {hu2019viznet}
\bibfield{author}{\bibinfo{person}{Kevin Hu}, \bibinfo{person}{Neil Gaikwad},
  \bibinfo{person}{Michiel Bakker}, \bibinfo{person}{Madelon Hulsebos},
  \bibinfo{person}{Emanuel Zgraggen}, \bibinfo{person}{C\'{e}sar Hidalgo},
  \bibinfo{person}{Tim Kraska}, \bibinfo{person}{Guoliang Li},
  \bibinfo{person}{Arvind Satyanarayan}, {and}
  \bibinfo{person}{{\c{C}}a{\u{g}}atay Demiralp}.}
  \bibinfo{year}{2019}\natexlab{}.
\newblock \showarticletitle{Viznet: Towards a large-scale visualization
  learning and benchmarking repository}. In
  \bibinfo{booktitle}{\emph{Proceedings of the 2019 CHI Conference on Human
  Factors in Computing Systems}}. \bibinfo{address}{Glasgow, Scotland, UK},
  \bibinfo{pages}{1--12}.
\newblock
\urldef\tempurl%
\url{https://doi.org/10.1145/3290605.3300892}
\showDOI{\tempurl}


\bibitem[Hulsebos et~al\mbox{.}(2019)]%
        {hulsebos2019sherlock}
\bibfield{author}{\bibinfo{person}{Madelon Hulsebos}, \bibinfo{person}{Kevin
  Hu}, \bibinfo{person}{Michiel Bakker}, \bibinfo{person}{Emanuel Zgraggen},
  \bibinfo{person}{Arvind Satyanarayan}, \bibinfo{person}{Tim Kraska},
  \bibinfo{person}{{\c{C}}agatay Demiralp}, {and} \bibinfo{person}{C{\'e}sar
  Hidalgo}.} \bibinfo{year}{2019}\natexlab{}.
\newblock \showarticletitle{Sherlock: A deep learning approach to semantic data
  type detection}. In \bibinfo{booktitle}{\emph{Proceedings of the 25th ACM
  SIGKDD International Conference on Knowledge Discovery \& Data Mining}}.
  \bibinfo{address}{Anchorage, AL, USA}, \bibinfo{pages}{1500--1508}.
\newblock
\urldef\tempurl%
\url{https://doi.org/10.1145/3292500.3330993}
\showDOI{\tempurl}


\bibitem[Kandel et~al\mbox{.}(2011)]%
        {kandel2011wrangler}
\bibfield{author}{\bibinfo{person}{Sean Kandel}, \bibinfo{person}{Andreas
  Paepcke}, \bibinfo{person}{Joseph Hellerstein}, {and}
  \bibinfo{person}{Jeffrey Heer}.} \bibinfo{year}{2011}\natexlab{}.
\newblock \showarticletitle{Wrangler: Interactive visual specification of data
  transformation scripts}. In \bibinfo{booktitle}{\emph{Proceedings of the 29th
  SIGCHI Conference on Human Factors in Computing Systems}}.
  \bibinfo{address}{Vancouver, BC, Canada}, \bibinfo{pages}{3363--3372}.
\newblock
\urldef\tempurl%
\url{https://doi.org/10.1145/1978942.1979444}
\showDOI{\tempurl}


\bibitem[Kingma and Ba(2015)]%
        {adam}
\bibfield{author}{\bibinfo{person}{Diederik~P. Kingma} {and}
  \bibinfo{person}{Jimmy Ba}.} \bibinfo{year}{2015}\natexlab{}.
\newblock \showarticletitle{Adam: {A} Method for Stochastic Optimization}. In
  \bibinfo{booktitle}{\emph{3rd International Conference on Learning
  Representations, (ICLR)}}. \bibinfo{address}{San Diego, CA, USA}.
\newblock
\urldef\tempurl%
\url{http://arxiv.org/abs/1412.6980}
\showURL{%
\tempurl}


\bibitem[Kipf and Welling(2016)]%
        {gcn}
\bibfield{author}{\bibinfo{person}{Thomas~N. Kipf} {and} \bibinfo{person}{Max
  Welling}.} \bibinfo{year}{2016}\natexlab{}.
\newblock \showarticletitle{Semi-Supervised Classification with Graph
  Convolutional Networks}. In \bibinfo{booktitle}{\emph{5th International
  Conference on Learning Representations, (ICLR)}}.
\newblock
\urldef\tempurl%
\url{http://arxiv.org/abs/1609.02907}
\showURL{%
\tempurl}


\bibitem[Le and Mikolov(2014)]%
        {pmlr-v32-le14}
\bibfield{author}{\bibinfo{person}{Quoc Le} {and} \bibinfo{person}{Tomas
  Mikolov}.} \bibinfo{year}{2014}\natexlab{}.
\newblock \showarticletitle{Distributed Representations of Sentences and
  Documents}. In \bibinfo{booktitle}{\emph{Proceedings of the 31st
  International Conference on Machine Learning (PMLR)}},
  Vol.~\bibinfo{volume}{32}. \bibinfo{address}{Bejing, China},
  \bibinfo{pages}{1188--1196}.
\newblock
\urldef\tempurl%
\url{https://doi.org/10.5555/3044805.3045025}
\showDOI{\tempurl}


\bibitem[Mikolov et~al\mbox{.}(2013)]%
        {mikolov2013efficient}
\bibfield{author}{\bibinfo{person}{Tomas Mikolov}, \bibinfo{person}{Kai Chen},
  \bibinfo{person}{Greg Corrado}, {and} \bibinfo{person}{Jeffrey Dean}.}
  \bibinfo{year}{2013}\natexlab{}.
\newblock \showarticletitle{Efficient estimation of word representations in
  vector space}. In \bibinfo{booktitle}{\emph{ICLR Workshop Papers}}.
  \bibinfo{publisher}{International Conference on Learning Representations}.
\newblock
\urldef\tempurl%
\url{https://arxiv.org/abs/1301.3781}
\showURL{%
\tempurl}


\bibitem[Pennington et~al\mbox{.}(2014)]%
        {pennington2014glove}
\bibfield{author}{\bibinfo{person}{Jeffrey Pennington},
  \bibinfo{person}{Richard Socher}, {and} \bibinfo{person}{Christopher~D
  Manning}.} \bibinfo{year}{2014}\natexlab{}.
\newblock \showarticletitle{Glove: Global vectors for word representation}. In
  \bibinfo{booktitle}{\emph{Proceedings of the 2014 Conference on Empirical
  Methods in Natural Language Processing (EMNLP)}}. \bibinfo{address}{Doha,
  Qatar}, \bibinfo{pages}{1532--1543}.
\newblock
\urldef\tempurl%
\url{https://doi.org/10.3115/v1/D14-1162}
\showDOI{\tempurl}


\bibitem[Rahm and Bernstein(2001)]%
        {rahm2001survey}
\bibfield{author}{\bibinfo{person}{Erhard Rahm} {and} \bibinfo{person}{Philip~A
  Bernstein}.} \bibinfo{year}{2001}\natexlab{}.
\newblock \showarticletitle{A survey of approaches to automatic schema
  matching}.
\newblock \bibinfo{journal}{\emph{the VLDB Journal}} \bibinfo{volume}{10},
  \bibinfo{number}{4} (\bibinfo{year}{2001}), \bibinfo{pages}{334--350}.
\newblock
\urldef\tempurl%
\url{https://doi.org/10.1007/s007780100057}
\showDOI{\tempurl}


\bibitem[Raman and Hellerstein(2001)]%
        {raman2001potter}
\bibfield{author}{\bibinfo{person}{Vijayshankar Raman} {and}
  \bibinfo{person}{Joseph~M Hellerstein}.} \bibinfo{year}{2001}\natexlab{}.
\newblock \showarticletitle{Potter's wheel: An interactive data cleaning
  system}. In \bibinfo{booktitle}{\emph{Proceedings of the 27th International
  Conference on Very Large Data Bases (VLDB)}}, Vol.~\bibinfo{volume}{1}.
  \bibinfo{address}{Roma, Italy}, \bibinfo{pages}{381--390}.
\newblock
\urldef\tempurl%
\url{https://doi.org/10.5555/645927.672045}
\showDOI{\tempurl}


\bibitem[Simonovsky and Komodakis(2017)]%
        {eccv}
\bibfield{author}{\bibinfo{person}{Martin Simonovsky} {and}
  \bibinfo{person}{Nikos Komodakis}.} \bibinfo{year}{2017}\natexlab{}.
\newblock \showarticletitle{Dynamic Edge-Conditioned Filters in Convolutional
  Neural Networks on Graphs}. In \bibinfo{booktitle}{\emph{2017 IEEE Conference
  on Computer Vision and Pattern Recognition (CVPR)}}.
  \bibinfo{address}{Hawaii, US}, \bibinfo{pages}{29--38}.
\newblock
\urldef\tempurl%
\url{http://arxiv.org/abs/1704.02901}
\showURL{%
\tempurl}


\bibitem[Vaswani et~al\mbox{.}(2017)]%
        {transformer}
\bibfield{author}{\bibinfo{person}{Ashish Vaswani}, \bibinfo{person}{Noam
  Shazeer}, \bibinfo{person}{Niki Parmar}, \bibinfo{person}{Jakob Uszkoreit},
  \bibinfo{person}{Llion Jones}, \bibinfo{person}{Aidan~N Gomez},
  \bibinfo{person}{\L~ukasz Kaiser}, {and} \bibinfo{person}{Illia Polosukhin}.}
  \bibinfo{year}{2017}\natexlab{}.
\newblock \showarticletitle{Attention is All you Need}. In
  \bibinfo{booktitle}{\emph{Proceedings of the 31st International Conference on
  Neural Information Processing Systems}}, Vol.~\bibinfo{volume}{30}.
  \bibinfo{address}{Long Beach, CA, US}.
\newblock
\urldef\tempurl%
\url{https://dl.acm.org/doi/10.5555/3295222.3295349}
\showURL{%
\tempurl}


\bibitem[Venetis et~al\mbox{.}(2011)]%
        {mind}
\bibfield{author}{\bibinfo{person}{Petros Venetis}, \bibinfo{person}{Alon
  Halevy}, \bibinfo{person}{Jayant Madhavan}, \bibinfo{person}{Marius
  Pa\c{s}ca}, \bibinfo{person}{Warren Shen}, \bibinfo{person}{Fei Wu},
  \bibinfo{person}{Gengxin Miao}, {and} \bibinfo{person}{Chung Wu}.}
  \bibinfo{year}{2011}\natexlab{}.
\newblock \showarticletitle{Recovering Semantics of Tables on the Web}. In
  \bibinfo{booktitle}{\emph{Proceedings of the 37th International Conference on
  Very Large Data Bases (VLDB)}}, Vol.~\bibinfo{volume}{4}.
  \bibinfo{address}{Seattle, WA, US}, \bibinfo{pages}{528–538}.
\newblock
\urldef\tempurl%
\url{https://doi.org/10.14778/2002938.2002939}
\showDOI{\tempurl}


\bibitem[Zhang et~al\mbox{.}(2020)]%
        {sato}
\bibfield{author}{\bibinfo{person}{Dan Zhang}, \bibinfo{person}{Yoshihiko
  Suhara}, \bibinfo{person}{Jinfeng Li}, \bibinfo{person}{Madelon Hulsebos},
  \bibinfo{person}{{\c{C}}a{\u{g}}atay Demiralp}, {and}
  \bibinfo{person}{Wang-Chiew Tan}.} \bibinfo{year}{2020}\natexlab{}.
\newblock \showarticletitle{Sato: Contextual semantic type detection in
  tables}. In \bibinfo{booktitle}{\emph{Proceedings of the 46th International
  Conference on Very Large Data Bases (VLDB)}}, Vol.~\bibinfo{volume}{13}.
  \bibinfo{address}{Tokyo, Japan}, \bibinfo{pages}{1835–1848}.
\newblock
\urldef\tempurl%
\url{https://doi.org/10.14778/3407790.3407793}
\showDOI{\tempurl}


\bibitem[Zheng et~al\mbox{.}(2022)]%
        {bytegnn}
\bibfield{author}{\bibinfo{person}{Chenguang Zheng}, \bibinfo{person}{Hongzhi
  Chen}, \bibinfo{person}{Yuxuan Cheng}, \bibinfo{person}{Zhezheng Song},
  \bibinfo{person}{Yifan Wu}, \bibinfo{person}{Changji Li},
  \bibinfo{person}{James Cheng}, \bibinfo{person}{Hao Yang}, {and}
  \bibinfo{person}{Shuai Zhang}.} \bibinfo{year}{2022}\natexlab{}.
\newblock \showarticletitle{ByteGNN: Efficient Graph Neural Network Training at
  Large Scale}. In \bibinfo{booktitle}{\emph{Proceedings of the 48th
  International Conference on Very Large Data Bases (VLDB)}},
  Vol.~\bibinfo{volume}{15}. \bibinfo{address}{Sydney, Australia},
  \bibinfo{pages}{1228–1242}.
\newblock
\showISSN{2150-8097}
\urldef\tempurl%
\url{https://doi.org/10.14778/3514061.3514069}
\showURL{%
\tempurl}


\end{thebibliography}


\end{document}